\documentclass[sigconf]{acmart}
\usepackage{algorithm}
\usepackage{algpseudocode}
\usepackage{tikz}
\usetikzlibrary{trees}
\usepackage{caption}
\usepackage{subcaption}

\newtheorem{definition}{Definition}
\newtheorem{theorem}{Theorem}

\AtBeginDocument{%
  \providecommand\BibTeX{{%
    \normalfont B\kern-0.5em{\scshape i\kern-0.25em b}\kern-0.8em\TeX}}}

\copyrightyear{2021}
\acmYear{2021}
\setcopyright{acmcopyright}\acmConference[CIKM '21]{Proceedings of the 30th ACM
International Conference on Information and Knowledge Management}{November
1--5, 2021}{Virtual Event, QLD, Australia}
\acmBooktitle{Proceedings of the 30th ACM International Conference on Information
and Knowledge Management (CIKM '21), November 1--5, 2021, Virtual Event, QLD,
Australia}
\acmPrice{15.00}
\acmDOI{10.1145/3459637.3482467}
\acmISBN{978-1-4503-8446-9/21/11}

\begin{document}
\sloppy
\fancyhead{}
\title{Random Sampling Plus Fake Data: Multidimensional Frequency Estimates With Local Differential Privacy}


\author{Héber H. Arcolezi}
\email{heber.hwang\_arcolezi@univ-fcomte.fr}
\orcid{0000-0001-8059-7094}
\affiliation{
  \institution{Femto-ST Institute, Univ. Bourg. Franche-Comt\'e, CNRS}
  \city{Belfort}
  \country{France}
%   \postcode{90000}
}

\author{Jean-François Couchot}
\email{jean-francois.couchot@univ-fcomte.fr}
\orcid{0000-0001-6437-5598}
\affiliation{
  \institution{Femto-ST Institute, Univ. Bourg. Franche-Comt\'e, CNRS}
  \city{Belfort}
  \country{France}
%   \postcode{90000}
}

\author{Bechara Al Bouna}
\email{bechara.albouna@ua.edu.lb}
\orcid{0000-0002-7741-9905}
\affiliation{
  \institution{TICKET Lab., Antonine University}
  \city{Hadat-Baabda}
  \country{Lebanon}
%   \postcode{90000}
}

\author{Xiaokui Xiao}
\email{xkxiao@nus.edu.sg}
\orcid{0000-0003-0914-4580}
\affiliation{
  \institution{School of Computing, National University of Singapore}
  \city{Singapore}
  \country{Singapore}
%   \postcode{90000}
}

\renewcommand{\shortauthors}{Arcolezi, et al.}%.

\begin{abstract}
  With local differential privacy (LDP), users can privatize their data and thus guarantee privacy properties before transmitting it to the server (a.k.a. the aggregator). One primary objective of LDP is frequency (or histogram) estimation, in which the aggregator estimates the number of users for each possible value. In practice, when a study with rich content on a population is desired, the interest is in the multiple attributes of the population, that is to say, in multidimensional data ($d \geq 2$). However, contrary to the problem of frequency estimation of a single attribute (the majority of the works), the multidimensional aspect imposes to pay particular attention to the privacy budget. This one can indeed grow extremely quickly due to the composition theorem. To the authors' knowledge, two solutions seem to stand out for this task: 1) splitting the privacy budget for each attribute, i.e., send each value with $\frac{\epsilon}{d}$-LDP (\textit{Spl}), and 2) random sampling a single attribute and spend all the privacy budget to send it with $\epsilon$-LDP (\textit{Smp}). Although \textit{Smp} adds additional sampling error, it has proven to provide higher data utility than the former \textit{Spl} solution. However, we argue that aggregators (who are also seen as attackers) are aware of the sampled attribute and its LDP value, which is protected by a "less strict" $e^{\epsilon}$ probability bound (rather than $e^{\epsilon/d}$). This way, we propose a solution named \underline{R}andom \underline{S}ampling plus \underline{F}ake \underline{D}ata (RS+FD), which allows creating \textit{uncertainty} over the sampled attribute by generating fake data for each non-sampled attribute; RS+FD further benefits from amplification by sampling. We theoretically and experimentally validate our proposed solution on both synthetic and real-world datasets to show that RS+FD achieves nearly the same or better utility than the state-of-the-art \textit{Smp} solution.
\end{abstract}





\maketitle

\section{Introduction} \label{sec:introduction}

\subsection{Background}

In recent years, differential privacy (DP)~\cite{Dwork2006,Dwork2006DP} has been increasingly accepted as the current standard for data privacy~\cite{dwork2014algorithmic,DL_DP,aktay2020google,linkedin}. In the centralized model of DP, a trusted curator has access to compute on the entire raw data of users (e.g., the Census Bureau~\cite{census,census2021}). By `trusted', we mean that curators do not misuse or leak private information from individuals. However, this assumption does not always hold in real life~\cite{data_breaches}. To address non-trusted services, with the local model of DP (LDP)~\cite{first_ldp}, each user applies a DP mechanism to their own data before sending it to an untrusted curator (a.k.a. the aggregator). The LDP model allows collecting data in unprecedented ways and, therefore, has led to several adoptions by industry. For instance, big tech companies like Google, Apple, and Microsoft, reported the implementation of LDP mechanisms to gather statistics in well-known systems (i.e., Google Chrome browser~\cite{rappor}, Apple iOS and macOS~\cite{apple}, and Windows 10 operation system~\cite{microsoft}). 

\subsection{Problem statement} \label{sub:problem_statement}
On collecting data, in practice, one is often interested in multiple attributes of a population, i.e., multidimensional data. For instance, in cloud services, demographic information (e.g., age, gender) and user habits could provide several insights to further develop solutions to specific groups. Similarly, in digital patient records, users might be linked with both their demographic and clinical information.

In this paper, we focus on the problem of private frequency (or histogram) estimation on multiple attributes with LDP. This is a primary objective of LDP, in which the data collector decodes all the privatized data of the users and can then estimate the number of users for each possible value. The single attribute frequency estimation task has received considerable attention in the literature~\cite{tianhao2017,kairouz2016discrete,Hadamard,Alvim2018,rappor,microsoft,Xiong2020,Zhao2019,Li2020} as it is a building block for more complex tasks (e.g., heavy hitter estimation~\cite{Bassily2015,Wang2021,bassily2017practical}, estimating marginals~\cite{Peng2019,Zhang2018,Ren2018,Fanti2016}, frequent itemset mining~\cite{Wang2018,Qin2016}).

In the LDP setting, the aggregator already knows the users' identifiers, but not their private data. We assume there are $d$ attributes $A=\{A_1,A_2,...,A_d\}$, where each attribute $A_j$ with a discrete domain $\mathcal{D}_j$ has a specific number of values $|A_j|=k_j$. Each user $u_i$ for $i \in \{1,2,...,n\}$ has a tuple $\textbf{v}^{(i)}=(v^{(i)}_{1},v^{(i)}_{2},...,v^{(i)}_{d})$, where $v^{(i)}_{j}$ represents the value of attribute $A_j$ in record $\textbf{v}^{(i)}$. Thus, for each attribute $A_j$, the analyzer's goal is to estimate a $k_j$-bins histogram, including the frequency of all values in $\mathcal{D}_j$.

\subsection{Context of the problem} \label{sub:problematic}

Regarding multiple attributes, as also noticed in the recent survey work on LDP in~\cite{Xiong2020}, most studies for collecting multidimensional data with LDP mainly focused on numerical data (e.g.,~\cite{xiao2,wang2019,Duchi2018,Wang2021_b}). Unlike the single attribute frequency estimation problem (the majority of the works), the multidimensional setting needs to consider the allocation of the privacy budget. To the authors' knowledge, there are mainly two solutions for satisfying LDP by randomizing $\textbf{v}$. We will simply omit the index notation $\textbf{v}^{(i)}$ in the analysis as we focus on one arbitrary user $u_i$ here. On the one hand, due to the composition theorem~\cite{dwork2014algorithmic}, users can split the privacy budget for each attribute and send all randomized values with $\frac{\epsilon}{d}$-LDP to the aggregator (\textit{Spl}). The other solution is based on random sampling a single attribute and spend all the privacy budget to send it (\textit{Smp}). More precisely, each user tells the aggregator which attribute is sampled, and what is the perturbed value for it ensuring $\epsilon$-LDP; the aggregator would not receive any information about the remaining $d-1$ attributes. 

Although the later \textit{Smp} solution adds sampling error, in the literature~\cite{wang2019,xiao2,tianhao2017,Arcolezi2021,Wang2021_b}, it has proven to provide higher data utility than the former \textit{Spl} solution. However, aggregators (who are also seen as attackers) are aware of the sampled attribute and its LDP value, which is protected by a "less strict" $e^{\epsilon}$ probability bound (rather than $e^{\epsilon/d}$). In other words, while both solutions provide $\epsilon$-LDP, we argue that using the \textit{Smp} solution may be unfair with some users. For instance, on collecting multidimensional health records (i.e., demographic and clinical data), users that randomly sample a demographic attribute (e.g., gender) might be less concerned to report their data than those whose sampled attribute is "disease" (e.g., if positive for human immunodeficiency viruses - HIV).

This way, there is a privacy-utility trade-off between the \textit{Spl} and \textit{Smp} solutions. With these elements in mind, we formulate the problematic of this paper as: \textit{For the same privacy budget $\epsilon$, is there a solution for multidimensional frequency estimates that provides better data utility than Spl and more protection than Smp?}

\subsection{Purpose and contributions}

In this paper, we intend to solve the aforementioned problematic by answering the following question: \textit{What if the sampling result (i.e., the selected attribute) was \textbf{not} disclosed with the aggregator?} Since the sampling step randomly selects an attribute $j \in [1,d]$ (we slightly abuse the notation and use $j$ for $A_j$), we propose that users add uncertainty about the sampled attribute through generating $d-1$ \textit{fake data}, i.e., one for each non-sampled attribute. 

\begin{figure}
    \centering
    \includegraphics[width=1\linewidth]{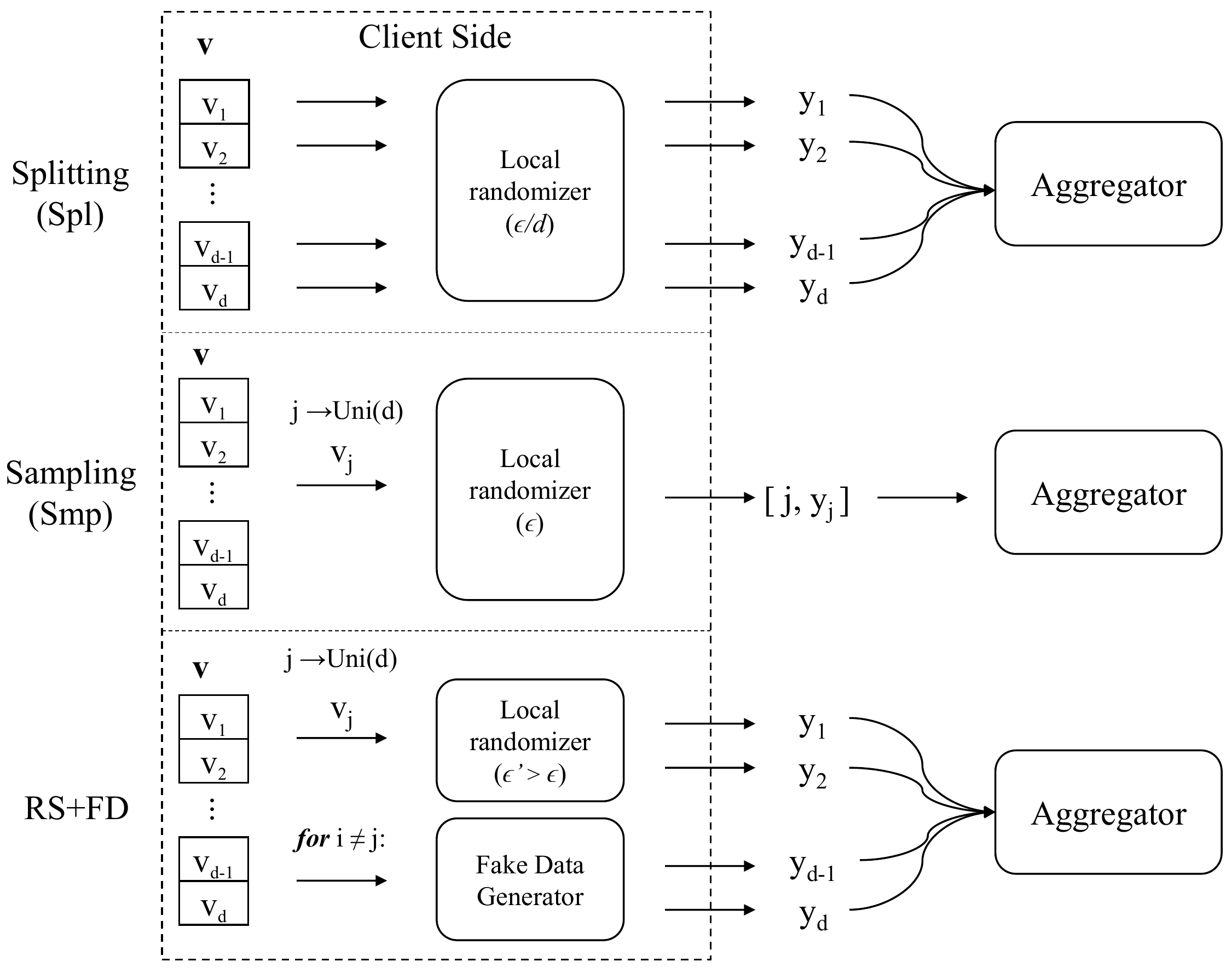}
    \caption{Overview of our random sampling plus fake data (RS+FD) solution in comparison with two known solutions, namely, \textit{Spl} and \textit{Smp}, where $Uni(d)=Uniform(\{1,2,...,d\})$.}
    \label{fig:system_overview}
\end{figure}

We call our solution \textit{\underline{R}andom \underline{S}ampling plus \underline{F}ake \underline{D}ata (RS+FD)}. Fig.~\ref{fig:system_overview} illustrates the overview of RS+FD in comparison with the aforementioned known solutions, namely, \textit{Spl} and \textit{Smp}. More precisely, with RS+FD, the client-side has two steps: local randomization and fake data generation. First, an LDP mechanism preserves privacy for the data of the sampled attribute. Second, the fake data generator provides fake data for each $d-1$ non-sampled attribute. This way, the privatized data is "hidden" among fake data and, hence, the sampling result is not disclosed along with the users' report (and statistics).

What is more, we notice that RS+FD can enjoy privacy amplification by sampling~\cite{Chaudhuri2006,Li2012,balle2018privacy,balle2020privacy,first_ldp}. That is, if one randomly sample a dataset without replacement using a sampling rate $\beta < 1$, it suffices to use a privacy budget $\epsilon' > \epsilon$ to satisfy $\epsilon$-DP, where $\frac{e^{\epsilon'}-1}{e^{\epsilon}-1} = \frac{1}{\beta}$~\cite{Li2012}. This way, given that the sampled dataset for each attribute has non-overlapping users, i.e., each user selects an attribute with sampling probability $\beta=\frac{1}{d}$, to satisfy $\epsilon$-LDP, each user can apply an LDP mechanism with $\epsilon'=\ln{\left( d \cdot (e^{\epsilon} - 1) + 1 \right)} \geq \epsilon$.

To summarize, this paper makes the following contributions:

\begin{itemize}
    \item We propose a novel solution, namely, RS+FD for multidimensional frequency estimates under LDP, which is generic to be used with any existing LDP mechanism developed for single-frequency estimation.
    
    \item Using state-of-the-art LDP mechanisms, we develop four protocols within RS+FD and analyze them analytically and experimentally.
    
    \item We conducted a comprehensive and extensive set of experiments on both synthetic and real-world datasets. Under the same privacy guarantee, results show that our proposed protocols with RS+FD achieve similar or better utility than using the state-of-the-art \textit{Smp} solution.

\end{itemize}

\noindent \textbf{Paper's outline:} The remainder of this paper is organized as follows. In Section~\ref{sec:background}, we revise the privacy notions that we are considering, i.e., LDP, the LDP mechanisms we further analyze in this paper, and amplification by sampling. In Section~\ref{sec:prop_metho}, we introduce our RS+FD solution, the integration of state-of-the-art LDP mechanisms within RS+FD, and their analysis. In Section~\ref{sec:results}, we present experimental results. In Section~\ref{sec:discussion}, we discuss our results and review related work. Lastly, in Section~\ref{sec:conclusion}, we present the concluding remarks and future directions.

\section{Preliminaries} \label{sec:background}

In this section, we briefly recall LDP (Subsection \ref{sub:ldp}), the LDP mechanisms we will apply in this paper (Subsection \ref{sub:grr_oue}), and amplification by sampling (Subsection \ref{sub:amp_sampling}).

\subsection{Local differential privacy} \label{sub:ldp}

Local differential privacy, initially formalized in~\cite{first_ldp}, protects an individual's privacy during the data collection process. A formal definition of LDP is given in the following:

\begin{definition}[$\epsilon$-Local Differential Privacy]\label{def:ldp} A randomized algorithm ${\mathcal {A}}$ satisfies $\epsilon$-LDP if, for any pair of input values $v_1, v_2 \in Domain(\mathcal {A})$ and any possible output $y$ of ${\mathcal {A}}$:

\begin{equation*}
    \Pr[{\mathcal {A}}(v_1) = y]\leq e^{\epsilon }\cdot \Pr[{\mathcal {A}}(v_2) = y] \textrm{.}
\label{eq:ldp}
\end{equation*}

\end{definition}

Similar to the centralized model of DP, LDP also enjoys several important properties, e.g., immunity to post-processing  ($F(\mathcal {A})$ is $\epsilon$-LDP for any function $F$) and composability~\cite{dwork2014algorithmic}. That is, combining the results from $m$ differentially private mechanisms also satisfies DP. If these mechanisms are applied separately in disjointed subsets of the dataset, $\epsilon = max(\epsilon_1$-, \ldots,  $\epsilon_m)$-LDP (parallel composition). On the other hand, if these mechanisms are sequentially applied to the same dataset, $\epsilon = \sum_{i=1}^{m}\epsilon_i$-LDP (sequential composition).

\subsection{LDP mechanisms} \label{sub:grr_oue}

Randomized response (RR), a surveying technique proposed by Warner~\cite{Warner1965}, has been the building block for many LDP mechanisms. Let $A_j=\{v_1,v_2,...,v_{k_j}\}$ be a set of $k_j=|A_j|$ values of a given attribute and let $\epsilon$ be the privacy budget, we review two state-of-the-art LDP mechanisms for single-frequency estimation (a.k.a. frequency oracles) that will be used in this paper.

\subsubsection{Generalized randomized response (GRR)} \label{sub:grr}

The \textit{k}-Ary RR~\cite{kairouz2016discrete} mechanism extends RR to the case of $k_j \geq 2$ and is also referred to as direct encoding~\cite{tianhao2017} or generalized RR (GRR)~\cite{Wang2018,Wang2020_post_process,Zhang2018}. Throughout this paper, we use the term GRR for this LDP mechanism. Given a value $B=v_i$, GRR($v_i$) outputs the true value $v_i$ with probability $p=\frac{e^{\epsilon}}{e^{\epsilon}+k_j-1}$, and any other value $v_l$ for $l \neq i$ with probability $q=\frac{1-p}{k_j-1}=\frac{1}{e^{\epsilon}+k_j-1}$. GRR satisfies $\epsilon$-LDP since $\frac{p}{q}=e^{\epsilon}$. 

The estimated frequency $\hat{f}(v_i)$ that a value $v_i$ occurs, for $i \in [1,k_j]$, is calculated as~\cite{tianhao2017,Wang2018}:

\begin{equation}\label{eq:est_pure}
    \hat{f}(v_i) = \frac{N_i - nq}{n(p - q)} \textrm{,}
\end{equation}

\noindent in which $N_i$ is the number of times the value $v_i$ has been reported and $n$ is the total number of users. In~\cite{tianhao2017}, it is shown that this is an unbiased estimation of the true frequency, and the variance of this estimation is $Var[\hat{f}(v_i)]= \frac{q(1-q)}{n(p-q)^2} + \frac{f(v_i)(1-p-q)}{n(p-q)}$. In the case of small $f(v_i) \sim 0$, this variance is dominated by the first term. Thus, the \textit{approximate variance} of this estimation for GRR is~\cite{tianhao2017}:

\begin{equation}\label{eq:var_grr}
    Var[\hat{f}_{GRR}(v_i)] = \frac{e^{\epsilon} + k_j - 2}{n(e^{\epsilon}-1)^2} \textrm{.}
\end{equation}

\subsubsection{Optimized unary encoding (OUE)} \label{sub:oue}

For a given value $v$, $B=Encode(v)$, where $B=[0,0,...,1,0,...0]$, a $k_j$-bit array in which only the $v$-th position is set to one. Subsequently, the bits from $B$ are flipped, depending on two parameters $p$ and $q$, to generate a privatized vector $B'$. More precisely, $Pr[B'[i]=1] =$ \textit{p} if $B[i]=1$ and $Pr[B'[i]=1] =$ \textit{q} if $B[i]=0$. This unary-encoding (UE) mechanism satisfies $\epsilon$-LDP for $\epsilon = ln\left( \frac{p(1-q)}{(1-p)q} \right )$~\cite{rappor}. Wang et al.~\cite{tianhao2017} propose optimized UE (OUE), which selects "optimized" parameters ($p=\frac{1}{2}$ and $q=\frac{1}{e^{\epsilon}+1}$) to minimize the \textit{approximate variance} of UE-based mechanisms while still satisfying $\epsilon$-LDP. The estimation method used in~\eqref{eq:est_pure} equally applies to OUE. As shown in~\cite{tianhao2017}, the OUE \textit{approximate variance} is calculated as:

\begin{equation}\label{eq:var_oue}
   Var[\hat{f}_{OUE}(v_i)] = \frac{4e^{\epsilon}}{n(e^{\epsilon}-1)^2} \textrm{.}
\end{equation}

\subsubsection{Adaptive LDP mechanism} \label{sub:adp}

Comparing~\eqref{eq:var_grr} with~\eqref{eq:var_oue}, elements $k_j-2+e^{\epsilon}$ is replaced by $4e^{\epsilon}$. Thus, as highlighted in~\cite{tianhao2017}, when $k_j < 3e^{\epsilon} +2$, the utility loss with GRR is lower than the one of OUE. Throughout this paper, we will use the term adaptive (ADP) to denote this best-effort and dynamic selection of LDP mechanism.

\subsection{Privacy amplification by sampling} \label{sub:amp_sampling}

One well-known approach for increasing the privacy of a DP mechanism is to apply the mechanism to a random subsample of the dataset~\cite{Chaudhuri2006,Li2012,balle2018privacy,balle2020privacy,first_ldp}. The intuition is that an attacker is unable to distinguish which data samples were used in the analysis. Li et al.~\cite[Theorem 1]{Li2012} theoretically prove this effect. 

\begin{theorem} \textbf{Amplification by Sampling}~\cite{Li2012}. Let $\mathcal{A}$ be an $\epsilon'$-DP mechanism and $\mathcal{S}$ to be a sampling algorithm with sampling rate $\beta$. Then, if $\mathcal{S}$ is first applied to a dataset $\mathbb{D}$, which is later privatized with $\mathcal{A}$, the derived result satisfies DP with $\epsilon=\ln{\left( 1 + \beta (e^{\epsilon'} + 1)  \right)}$.

\end{theorem}

\section{Random Sampling Plus Fake Data} \label{sec:prop_metho}

In this section, we present the overview of our RS+FD solution (Subsection~\ref{sub:overview}), and the integration of the local randomizers presented in Subsection~\ref{sub:grr_oue} within RS+FD (Subsections~\ref{sub:rs+fd_grr},~\ref{sub:rs+fd_oue}, and~\ref{sub:rs+fd_adp}).

\subsection{Overview of RS+FD} \label{sub:overview}

We consider the local DP model, in which there are two entities, namely, users and the aggregator (an untrusted curator). Let $n$ be the total number of users, $d$ be the total number of attributes, $\textbf{k}=[k_1,k_2,...,k_d]$ be the domain size of each attribute, $\mathcal{A}$ be a local randomizer, and $\epsilon$ be the privacy budget. Each user holds a tuple $\textbf{v}=(v_1,v_2,...,v_d)$, i.e., a private value per attribute.

\noindent \textbf{Client-Side.} The client-side is split into two steps, namely, local randomization and fake data generation (cf. Fig.~\ref{fig:system_overview}). Initially, each user samples a unique attribute $j$ uniformly at random and applies an LDP mechanism to its value $v_j$. Indeed, RS+FD is generic to be applied with any existing LDP mechanisms (e.g., GRR~\cite{kairouz2016discrete}, UE-based protocols~\cite{rappor,tianhao2017}, Hadamard Response~\cite{Hadamard}). Next, for each $d-1$ non-sampled attribute $i$, the user generates one random fake data. Finally, each user sends the (LDP or fake) value of each attribute to the aggregator, i.e., a tuple $\textbf{y}=(y_1,y_2,...,y_d)$. This way, the sampling result is not disclosed with the aggregator. In summary, Alg.~\ref{alg:rs+fd} exhibits the pseudocode of our RS+FD solution.

\noindent \textbf{Aggregator.} For each attribute $j \in [1,d]$, the aggregator performs frequency (or histogram) estimation on the collected data by removing bias introduced by the local randomizer and fake data.

\begin{algorithm}

\caption{\underline{R}andom \underline{S}ampling plus \underline{F}ake \underline{D}ata (RS+FD)}
\label{alg:rs+fd}
\begin{algorithmic}[1]
\scriptsize
\Statex \textbf{Input :} tuple $\textbf{v} = (v_1,v_2,..., v_d)$, domain size of attributes $\textbf{k}=[k_1,k_2,...,k_d]$, privacy parameter $\epsilon$, local randomizer $\mathcal{A}$. 
\Statex \textbf{Output :} privatized tuple $\textbf{y}=(y_1,y_2,...,y_d)$.

\State $\epsilon' = \ln{\left( d \cdot (e^{\epsilon} - 1) + 1 \right)}$ \Comment{amplification by sampling~\cite{Li2012}} 

\State $j \gets Uniform(\{1,2,...,d \})$ \Comment{Selection of attribute to privatize}

\State $B_j \gets v_j$

\State $y_j \gets \mathcal{A}(B_j, k_j, \epsilon')$ \Comment{privatize data of the sampled attribute}

\State \textbf{for} $i \in \{1,2,...,d\}/j$ \textbf{do}\Comment{non-sampled attributes} 

\State  \hskip1em $y_i \gets \textit{Uniform}(\{1,...,k_i\}) $ \Comment{generate fake data}

\State \textbf{end for}

\Statex \textbf{return :} $\textbf{y}=(y_1,y_2,...,y_d)$ \Comment{sampling result is not disclosed}
\end{algorithmic}
\end{algorithm}

\noindent \textbf{Privacy analysis.} Let $\mathcal{A}$ be any existing LDP mechanism, Algorithm~\ref{alg:rs+fd} satisfies $\epsilon$-LDP, in a way that $\epsilon'=\ln{\left( d \cdot (e^{\epsilon} - 1) + 1 \right)}$. Indeed, we observe that our scenario is equivalent to sampling a dataset $\mathbb{D}$ without replacement with sampling rate $\beta=\frac{1}{d}$ in the centralized setting of DP, which enjoys privacy amplification (cf. Subsection~\ref{sub:amp_sampling}). With the local model, users privatize their data locally with a DP model. This way, to satisfy $\epsilon$-LDP, an amplified privacy parameter $\epsilon' > \epsilon$ can be used.

\noindent \textbf{Limitations.} Similar to other sampling-based methods for collecting multidimensional data under LDP~\cite{Duchi2018,xiao2,wang2019,Wang2021_b}, our RS+FD solution also entails \textit{sampling error}, which is due to observing a sample instead of the entire population. In addition, in comparison with the \textit{Smp} solution, RS+FD requires more computation on the user side because of the fake data generation part. Yet, communication cost is still equal to the \textit{Spl} solution, i.e., each user sends one message per attribute. Lastly, while RS+FD utilizes an amplified $\epsilon' \geq \epsilon$, there is also bias generated from uniform fake data that may require a sufficient number of users $n$ to eliminate the noise.

\subsection{RS+FD with GRR} \label{sub:rs+fd_grr}

\noindent \textbf{Client side.} Integrating GRR as the local randomizer $\mathcal{A}$ into Alg.~\ref{alg:rs+fd} (RS+FD[GRR]) requires no modification. Initially, on the client-side, each user randomly samples an attribute $j$. Next, the value $v_j$ is privatized with GRR (cf. Subsection~\ref{sub:grr}) using the size of the domain $k_j$ and the privacy parameter $\epsilon'$. In addition, for each non-sampled $d-1$ attribute $i$, the user also generates fake data uniformly at random according to the domain size $k_i$. Lastly, the user transmits the privatize tuple $\textbf{y}$, which includes the LDP value of the true data "hidden" among fake data.

\noindent \textbf{Aggregator RS+FD[GRR].} On the server-side, for each attribute $j\in[1,d]$, the aggregator estimates $\hat{f}(v_i)$ for the frequency of each value $i \in [1,k_j]$ as:

\begin{equation}\label{eq:est_rs+fd_grr}
    \hat{f}(v_i) = \frac{N_i dk_j - n(d - 1 + qk_j)}{nk_j(p-q)} \textrm{,} 
\end{equation}

\noindent in which $N_i$ is the number of times the value $v_i$ has been reported, $p=\frac{e^{\epsilon'}}{e^{\epsilon'} + k_j - 1}$, and $q=\frac{1-p}{k_j-1}$.

\begin{theorem} \label{theo:est_grr} For $j\in[1,d]$, the estimation result $\hat{f}(v_i)$ in~\eqref{eq:est_rs+fd_grr} is an unbiased estimation of $f (v_i)$ for any value $v_i \in \mathcal{D}_j$.
\end{theorem}

\noindent \textit{Proof~\ref{theo:est_grr}}

\begin{equation*}
\begin{aligned}
    E[\hat{f}(v_i)] &= E\left[ \frac{N_i d k_j - n(d - 1 + q k_j)}{n k_j (p-q)} \right] \\
    &= \frac{d }{n(p-q)} E[Ni] -  \frac{ d - 1 + q k_j}{k_j (p-q)}  \textrm{.}
\end{aligned}
\end{equation*}

Let us focus on 

\begin{equation*}
\begin{aligned}
    E[N_i] &= \frac{1}{d} \left( p n f (v_i) + q (n - n f (v_i))\right)  + \frac{d-1}{d k_j} n\\
    &= \frac{n}{d} \left(f (v_i)(p-q) + q   + \frac{d-1}{k_j} \right)   \textrm{.}
\end{aligned}
\end{equation*}

Thus,

\begin{equation*}
    E[\hat{f}(v_i)] = f(v_i) \textrm{.}
\end{equation*}

\begin{theorem} \label{theo:variance_grr} The variance of the estimation in~\eqref{eq:est_rs+fd_grr} is:

\begin{equation}\label{var:rs+fd_grr}
\begin{gathered}
    \operatorname{VAR}(\hat{f}(v_i)) = \frac{d^2 \delta (1-\delta)}{n (p-q)^2} \textrm{, where} \\
    \delta = \frac{1}{d} \left( q + f(v_i) (p-q) + \frac{(d-1)}{k_j} \right ) \textrm{.}
\end{gathered}
\end{equation}

\end{theorem}

\noindent \textit{Proof~\ref{theo:variance_grr}}

Thanks to~\eqref{eq:est_rs+fd_grr} we have

\begin{equation}
\operatorname{VAR}(\hat{f}(v_i)) = 
\frac{\operatorname{VAR}(N_i) d^2}{n^2 (p-q)^2}  \textrm{.}
\end{equation}

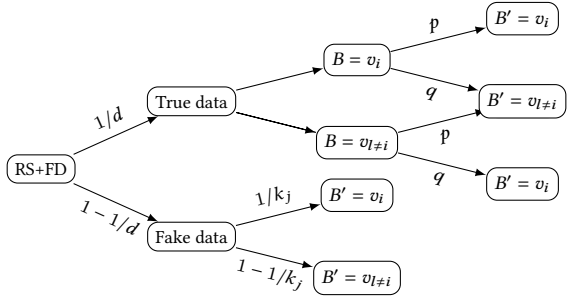
\begin{figure}[ht]
\centering
\tikzset{
  treenode/.style = {shape=rectangle, rounded corners,
                     draw,align=center,
                     top color=white},
  root/.style     = {treenode},
  env/.style      = {treenode},
  dummy/.style    = {circle,draw}
}
\tikzstyle{level 1}=[level distance=2cm, sibling distance=1.8cm]
\tikzstyle{level 2}=[level distance=2.2cm, sibling distance=1.1cm]

\begin{tikzpicture}
  [
    grow                    = right,
    edge from parent/.style = {draw, -latex},
    every node/.style       = {font=\footnotesize},
    sloped
  ]
  \node [root] {RS+FD}
    child { node [env] {Fake data}
        child { node [env] {$B'=v_{l\neq i}$}
          edge from parent node [below] {$1-1/k_j$} }
        child { node [env] {$B'=v_i$}
          edge from parent node [above] {$1/k_j$} }
        edge from parent node [below] {$1-1/d$} }
    child { node [env] {True data}
        child { node [env] {$B=v_{l\neq i}$}  
         child { node [env] {$B'=v_i$}
            edge from parent node [below] {$q$}}
            child { node [env] {$B'=v_i$}
            edge from parent node [below] {$p$}}
            edge from parent node [below] {}
            edge from parent node [below] {}}
        child { node [env] {$B=v_i$} 
            child { node [env] {$B'=v_{l\neq i}$}
            edge from parent node [below] {$q$}}
            child { node [env] {$B'=v_i$}
            edge from parent node [above] {$p$}}
            edge from parent node [above] {}}
        edge from parent node [above] {$1/d$}};
\end{tikzpicture}
\caption{Probability tree for Eq.~\eqref{eq:est_rs+fd_grr} (RS+FD[GRR]).} 
\label{fig:prob_tree_rsfd_grr}
\end{figure}

Since $N_i$ is the number of times value $v_i$ is observed, it can be defined as $N_i = \sum_{z=1}^n X_z$ where $X_z$ is equal to 1 if the user $z$, 
$1 \le z \le n$ reports value $v_i$, and 0 otherwise. We thus have 
$
\operatorname{VAR}(N_i) 
= \sum_{z=1}^n \operatorname{VAR}(X_z) 
= n \operatorname{VAR}(X)$, since all the users are independent. \textcolor{black}{According to the probability tree in Fig.~\ref{fig:prob_tree_rsfd_grr}, }
\[
P(X = 1) = P(X^2 = 1) = \delta = \frac{1}{d} \left( q + f(v_i) (p-q) + \frac{(d-1)}{k_j} \right )  \textrm{.}
\]
We thus have $\operatorname{VAR}(X)= \delta - \delta^2 = \delta(1 - \delta) $ and, finally,

\begin{equation} \label{var:generic}
\operatorname{VAR}(\hat{f}(v_i)) =
\frac{d^2 \delta (1-\delta)}{n (p-q)^2}.
\end{equation}

\subsection{RS+FD with OUE}\label{sub:rs+fd_oue}

\noindent \textbf{Client side.} To use UE-based protocols (OUE in our work) as local randomizer $\mathcal{A}$ in Alg.~\ref{alg:rs+fd}, there is, first, a need to define the fake data generation procedure. We propose two solutions: (i) RS+FD[OUE-z] in Alg.~\ref{alg:rs+fd_oue_z}, which applies OUE to $d-1$ zero-vectors, and (ii) RS+FD[OUE-r] in Alg.~\ref{alg:rs+fd_oue_r}, which applies OUE to $d-1$ one-hot-encoded fake data (uniform at random). We note that, on the one hand, introducing fake data through privatizing zero-vectors introduces less noise as there is only one parameter to perturb each bit, i.e., $Pr[0\rightarrow 1]=q$. On the other hand, starting with zero-vectors may not suffice to "hide" the sampled attribute if the perturbation probability $q$ is too small. Studying this effect is out of the scope of this paper and is left as future work. 

\begin{algorithm}[t]
\caption{RS+FD[OUE-z]}
\label{alg:rs+fd_oue_z}
\begin{algorithmic}[1]
\scriptsize
\Statex \textbf{Input :} tuple $\textbf{v} = (v_1,v_2,..., v_d)$, domain size of attributes $\textbf{k}=[k_1,k_2,...,k_d]$, privacy parameter $\epsilon$, local randomizer OUE. 
\Statex \textbf{Output :} privatized tuple $\textbf{B}'=(B_1',B_2',...,B_d')$.

\State $\epsilon' = \ln{\left( d \cdot (e^{\epsilon} - 1) + 1 \right)}$ \Comment{amplification by sampling~\cite{Li2012}} 

\State $j \gets Uniform(\{1,2,...,d \})$ \Comment{Selection of attribute to privatize}

\State $B_j=Encode(v_j)=[0,0,...,1,0,...0]$ \Comment{one-hot-encoding}

\State $B_j' \gets OUE(B_j, \epsilon')$ \Comment{privatize real data with OUE}

\State \textbf{for} $i \in \{1,2,...,d\}/j$ \textbf{do}\Comment{non-sampled attributes} 

\State  \hskip1em $B_i \gets [0,0,...,0] $ \Comment{initialize zero-vectors}

\State  \hskip1em $B_i' \gets OUE(B_i, \epsilon')$  \Comment{randomize zero-vector with OUE}

\State \textbf{end for}

\Statex \textbf{return :} $\textbf{B}'=(B_1',B_2',...,B_d')$ \Comment{sampling result is not disclosed}
\end{algorithmic}
\end{algorithm}

\begin{algorithm}[t]
\caption{RS+FD[OUE-r]}
\label{alg:rs+fd_oue_r}
\begin{algorithmic}[1]
\scriptsize
\Statex \textbf{Input :} tuple $\textbf{v} = (v_1,v_2,..., v_d)$, domain size of attributes $\textbf{k}=[k_1,k_2,...,k_d]$, privacy parameter $\epsilon$, local randomizer OUE. 
\Statex \textbf{Output :} privatized tuple $\textbf{B}'=(B_1',B_2',...,B_d')$.

\State $\epsilon' = \ln{\left( d \cdot (e^{\epsilon} - 1) + 1 \right)}$ \Comment{amplification by sampling~\cite{Li2012}} 

\State $j \gets Uniform(\{1,2,...,d \})$ \Comment{Selection of attribute to privatize}

\State $B_j=Encode(v_j)=[0,0,...,1,0,...0]$ \Comment{one-hot-encoding}

\State $B_j' \gets OUE(B_j, \epsilon')$ \Comment{privatize real data with OUE}

\State \textbf{for} $i \in \{1,2,...,d\}/j$ \textbf{do}\Comment{non-sampled attributes} 

\State  \hskip1em $y_i \gets \textit{Uniform}(\{1,...,k_i\}) $ \Comment{generate fake data}

\State  \hskip1em $B_i \gets Encode(y_i) $ \Comment{one-hot-encoding}

\State  \hskip1em $B_i' \gets OUE(B_i, \epsilon')$  \Comment{randomize fake data with OUE}

\State \textbf{end for}

\Statex \textbf{return :} $\textbf{B}'=(B_1',B_2',...,B_d')$ \Comment{sampling result is not disclosed}
\end{algorithmic}
\end{algorithm}

\noindent \textbf{Aggregator RS+FD[OUE-z].} On the server-side, if fake data are generated with OUE applied to zero-vectors, as in Alg.~\ref{alg:rs+fd_oue_z}, for each attribute $j\in[1,d]$, the aggregator estimates $\hat{f}(v_i)$ for the frequency of each value $i \in [1,k_j]$ as:

\begin{equation}\label{eq:est_rs+fd_oue_zeros}
    \hat{f}(v_i) =  \frac{d(N_i  - nq)}{n(p-q)}  \textrm{,}
\end{equation}

\noindent in which $N_i$ is the number of times the value $v_i$ has been reported, $n$ is the total number of users, $p=\frac{1}{2}$, and $q=\frac{1}{e^{\epsilon'}+1}$.

\begin{theorem} \label{theo:est_oue_z} For $j\in[1,d]$, the estimation result $\hat{f}(v_i)$ in~\eqref{eq:est_rs+fd_oue_zeros} is an unbiased estimation of $f (v_i)$ for any value $v_i \in \mathcal{D}_j$.
\end{theorem}

\noindent \textit{Proof~\ref{theo:est_oue_z}}

\begin{equation*}
\begin{aligned}
     E[\hat{f}(v_i)] &=  E\left[\frac{d(N_i  - nq)}{n(p-q)} \right] =   \frac{d(E[N_i]  - nq)}{n(p-q)} \\
     &= \frac{d}{n(p-q)}E[N_i]  - \dfrac{dq}{p-q}.
\end{aligned}
\end{equation*}

We have successively 

\begin{equation*}
\begin{aligned}
      E[N_i] &= \frac{n}{d} \left( p  f (v_i) + q (1 -  f (v_i))\right)  + \frac{(d-1)nq}{d}\\
     &= \frac{n}{d} \left( f(v_i)(p-q)   + dq \right) \textrm{.}
\end{aligned}
\end{equation*}

Thus,

\begin{equation*}
    E[\hat{f}(v_i)]  =  f(v_i)\textrm{.}
\end{equation*}

\begin{theorem} \label{theo:variance_oue_z} The variance of the estimation in~\eqref{eq:est_rs+fd_oue_zeros} is:

\begin{equation}\label{var:rs+fd_oue_z}
\begin{gathered}
    \operatorname{VAR}(\hat{f}(v_i)) = \frac{d^2 \delta (1-\delta)}{n (p-q)^2} \textrm{, where} \\
    \delta = \frac{1}{d} \left( dq + f(v_i) (p-q) \right) \textrm{.}
\end{gathered}
\end{equation}

\end{theorem}

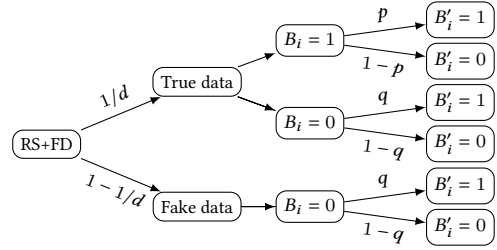
\begin{figure}[ht]
\centering
\tikzset{
  treenode/.style = {shape=rectangle, rounded corners,
                     draw,align=center,
                     top color=white},
  root/.style     = {treenode},
  env/.style      = {treenode},
  dummy/.style    = {circle,draw}
}
\tikzstyle{level 1}=[level distance=2cm, sibling distance=1.66cm]
\tikzstyle{level 2}=[level distance=1.5cm, sibling distance=1.1cm]
\tikzstyle{level 3}=[level distance=2cm, sibling distance=0.55cm]

\begin{tikzpicture}
  [
    grow                    = right,
    edge from parent/.style = {draw, -latex},
    every node/.style       = {font=\footnotesize},
    sloped
  ]
  \node [root] {RS+FD}
    child { node [env] {Fake data}
        child { node [env] {$B_i=0$}  
         child { node [env] {$B_i'=0$}
            edge from parent node [below] {$1-q$}}
            child { node [env] {$B_i'=1$}
            edge from parent node [above] {$q$}}
            edge from parent node [below] {}
            edge from parent node [below] {}} 
            edge from parent node [below] {$1-1/d$}}
    child { node [env] {True data}
        child { node [env] {$B_i=0$}  
         child { node [env] {$B_i'=0$}
            edge from parent node [below] {$1-q$}}
            child { node [env] {$B_i'=1$}
            edge from parent node [above] {$q$}}
            edge from parent node [below] {}
            edge from parent node [below] {}}
        child { node [env] {$B_i=1$} 
            child { node [env] {$B_i'=0$}
            edge from parent node [below] {$1-p$}}
            child { node [env] {$B_i'=1$}
            edge from parent node [above] {$p$}}
            edge from parent node [above] {}}
        edge from parent node [above] {$1/d$}};
\end{tikzpicture}
\caption{Probability tree for Eq.~\eqref{eq:est_rs+fd_oue_zeros} (RS+FD[OUE-z]).} 
\label{fig:prob_tree_rsfd_oue_z}
\end{figure}

The proof for Theorem~\ref{theo:variance_oue_z} follows \textit{Proof 3} and is omitted here. \textcolor{black}{In this case, $\delta$ follows the probability tree in Fig.~\ref{fig:prob_tree_rsfd_oue_z}}.

\noindent \textbf{Aggregator RS+FD[OUE-r].} Otherwise, if fake data are generated with OUE applied to one-hot-encoded random data, as in Alg.~\ref{alg:rs+fd_oue_r}, for each attribute $j\in[1,d]$, the aggregator estimates $\hat{f}(v_i)$ for the frequency of each value $i \in [1,k_j]$ as:

\begin{equation}\label{eq:est_rs+fd_oue_random}
    \hat{f}(v_i) =  \frac{N_i d k_j - n\left[ qk_j + (p-q)(d-1) + qk_j(d-1)) \right]}{nk_j(p-q)} \textrm{,} 
\end{equation}

\noindent in which $N_i$ is the number of times the value $v_i$ has been reported, $p=\frac{1}{2}$, and $q=\frac{1}{e^{\epsilon'}+1}$.

\begin{theorem} \label{theo:est_oue_r} For $j\in[1,d]$, the estimation result $\hat{f}(v_i)$ in~\eqref{eq:est_rs+fd_oue_random} is an unbiased estimation of $f (v_i)$ for any value $v_i \in \mathcal{D}_j$.
\end{theorem}

\noindent \textit{Proof~\ref{theo:est_oue_r}}

\begin{equation*}
\begin{aligned}
     E[\hat{f}(v_i)] &=  E\left[\frac{N_i d k_j - n\left[ qk_j + (p-q)(d-1) + qk_j(d-1)) \right]}{nk_j(p-q)}\right] \\
     &= \frac{d E[N_i]}{n(p-q)} - \frac{(p-q)(d-1) + qdk_j }{k_j(p-q)} .
\end{aligned}
\end{equation*}

We have successively 

\begin{equation*}
\begin{aligned}
     E[N_i] &= \frac{n}{d} \left( p  f (v_i) + q (1 -  f (v_i))\right)  + \frac{n(d-1)}{d}(\frac{p}{k_j} + \frac{k_j-1}{k_j}q)\\
     &=  \frac{n}{d} \left( f(v_i)(p-q) + q)\right)  + \frac{n(d-1)}{dk_j}(p-q  + k_jq) \textrm{.}
\end{aligned}
\end{equation*}

Thus,

\[
E[\hat{f}(v_i)]  =  f(v_i) \textrm{.}
\] 

\begin{theorem} \label{theo:variance_oue_r} The variance of the estimation in~\eqref{eq:est_rs+fd_oue_random} is:

\begin{equation}\label{var:rs+fd_oue_r}
\begin{gathered}
    \operatorname{VAR}(\hat{f}(v_i)) = \frac{d^2 \delta (1-\delta)}{n (p-q)^2} \textrm{, where} \\
    \delta = \frac{1}{d} \left(q + f(v_i) (p-q) + \frac{(d-1)}{k_j}\left( p + (k_j-1)q \right)\right) \textrm{.}
\end{gathered}
\end{equation}

\end{theorem}

\begin{figure}[ht]
\centering
\tikzset{
  treenode/.style = {shape=rectangle, rounded corners,
                     draw,align=center,
                     top color=white},
  root/.style     = {treenode},
  env/.style      = {treenode},
  dummy/.style    = {circle,draw}
}
\tikzstyle{level 1}=[level distance=2cm, sibling distance=2.2cm]
\tikzstyle{level 2}=[level distance=2.2cm, sibling distance=1.1cm]
\tikzstyle{level 3}=[level distance=2cm, sibling distance=0.55cm]

\begin{tikzpicture}
  [
    grow                    = right,
    edge from parent/.style = {draw, -latex},
    every node/.style       = {font=\footnotesize},
    sloped
  ]
  \node [root] {RS+FD}
    child { node [env] {Fake data}
        child { node [env] {$B_i=0$}  
         child { node [env] {$B_i'=0$}
            edge from parent node [below] {$1-q$}}
            child { node [env] {$B_i'=1$}
            edge from parent node [above] {$q$}}
            edge from parent node [below] {$1-1/k_j$}
            edge from parent node [below] {}}
        child { node [env] {$B_i=1$} 
            child { node [env] {$B_i'=0$}
            edge from parent node [below] {$1-p$}}
            child { node [env] {$B_i'=1$}
            edge from parent node [above] {$p$}}
            edge from parent node [above] {$1/k_j$}}
        edge from parent node [below] {$1-1/d$} }
    child { node [env] {True data}
        child { node [env] {$B_i=0$}  
         child { node [env] {$B_i'=0$}
            edge from parent node [below] {$1-q$}}
            child { node [env] {$B_i'=1$}
            edge from parent node [above] {$q$}}
            edge from parent node [below] {}
            edge from parent node [below] {}}
        child { node [env] {$B_i=1$} 
            child { node [env] {$B_i'=0$}
            edge from parent node [below] {$1-p$}}
            child { node [env] {$B_i'=1$}
            edge from parent node [above] {$p$}}
            edge from parent node [above] {}}
        edge from parent node [above] {$1/d$}};
\end{tikzpicture}
\caption{Probability tree for Eq.~\eqref{eq:est_rs+fd_oue_random} (RS+FD[OUE-r]).} 
\label{fig:prob_tree_rsfd_oue_r}
\end{figure}
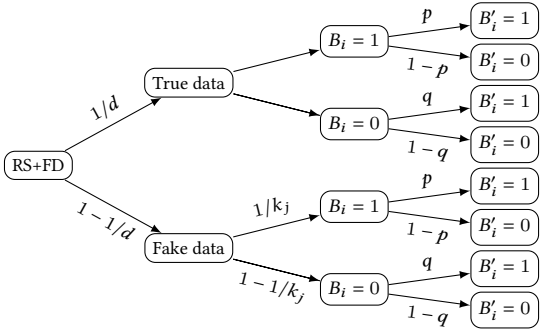

The proof for Theorem~\ref{theo:variance_oue_r} follows \textit{Proof 3} and is omitted here. \textcolor{black}{In this case, $\delta$ follows the probability tree in Fig.~\ref{fig:prob_tree_rsfd_oue_r}.}

\subsection{Analytical analysis: RS+FD with ADP} \label{sub:rs+fd_adp}

In a multidimensional setting with different domain sizes for each attribute, a dynamic selection of LDP mechanisms is preferred (as in Subsection~\ref{sub:adp}). Because our estimators in~\eqref{eq:est_rs+fd_grr},~\eqref{eq:est_rs+fd_oue_zeros}, and~\eqref{eq:est_rs+fd_oue_random} are unbiased, their variance is equal to the mean squared error (MSE) that is commonly used in practice as an accuracy metric~\cite{Wang2020,Wang2020_post_process,Wang2021_b}. 
In this work, we analyze the \textit{approximate variances} $\operatorname{VAR}_1$ for RS+FD[GRR] in~\eqref{var:rs+fd_grr} and $\operatorname{VAR}_2$ for RS+FD[OUE-z] in~\eqref{var:rs+fd_oue_z}, in which $f(v_i)=0$. This is, first, because under the local model of DP, the real frequency $f(v_i)$ is unknown and, second, because in real life the vast majority of values appear very infrequently. Notice that this is common practice in the literature (e.g., cf.~\cite{tianhao2017,Wang2021_b}), which provides an approximation for the variance of the protocols. 

Assume there are $d\geq 2$ attributes with domain size $\textbf{k}=[k_1,k_2,...,k_d]$ and a privacy budget $\epsilon'$. For each attribute $j$ with domain size $k_j$, to select RS+FD[GRR], we are then left to evaluate if $\operatorname{VAR}_1 \leq \operatorname{VAR}_2$. This is equivalent to check whether, 

\begin{equation}\label{ineq:variance}
\frac{d^2\delta_1 (1-\delta_1)}{n(p_1-q_1)^2} - \frac{d^2\delta_2 (1-\delta_2)}{n(p_2-q_2)^2} \leq 0 \textrm{,}
\end{equation}

\noindent in which
$p_1 = \frac{e^{\epsilon'}}{e^{\epsilon'} + k_j - 1}$, 
$q_1 = \frac{1-p_1}{k_j-1}$,
$p_2=\frac{1}{2}$, 
$q_2=\frac{1}{e^{\epsilon'}+1}$,
$\delta_1 = 
\frac{1}{d}
\left( 
q_1 + \frac{d-1}{k_j}
\right)$, and 
$
\delta_2 =q_2$. 
\textbf{In other words, if \eqref{ineq:variance} is less than or equal to zero, the utility loss is lower with RS+FD[GRR]; otherwise, if \eqref{ineq:variance} is positive, RS+FD[OUE-z] should be selected. Throughout this paper, we will refer to this dynamic selection of our protocols as RS+FD[ADP].}

\begin{figure}[htb]
    \centering
    \includegraphics[width=0.64\linewidth]{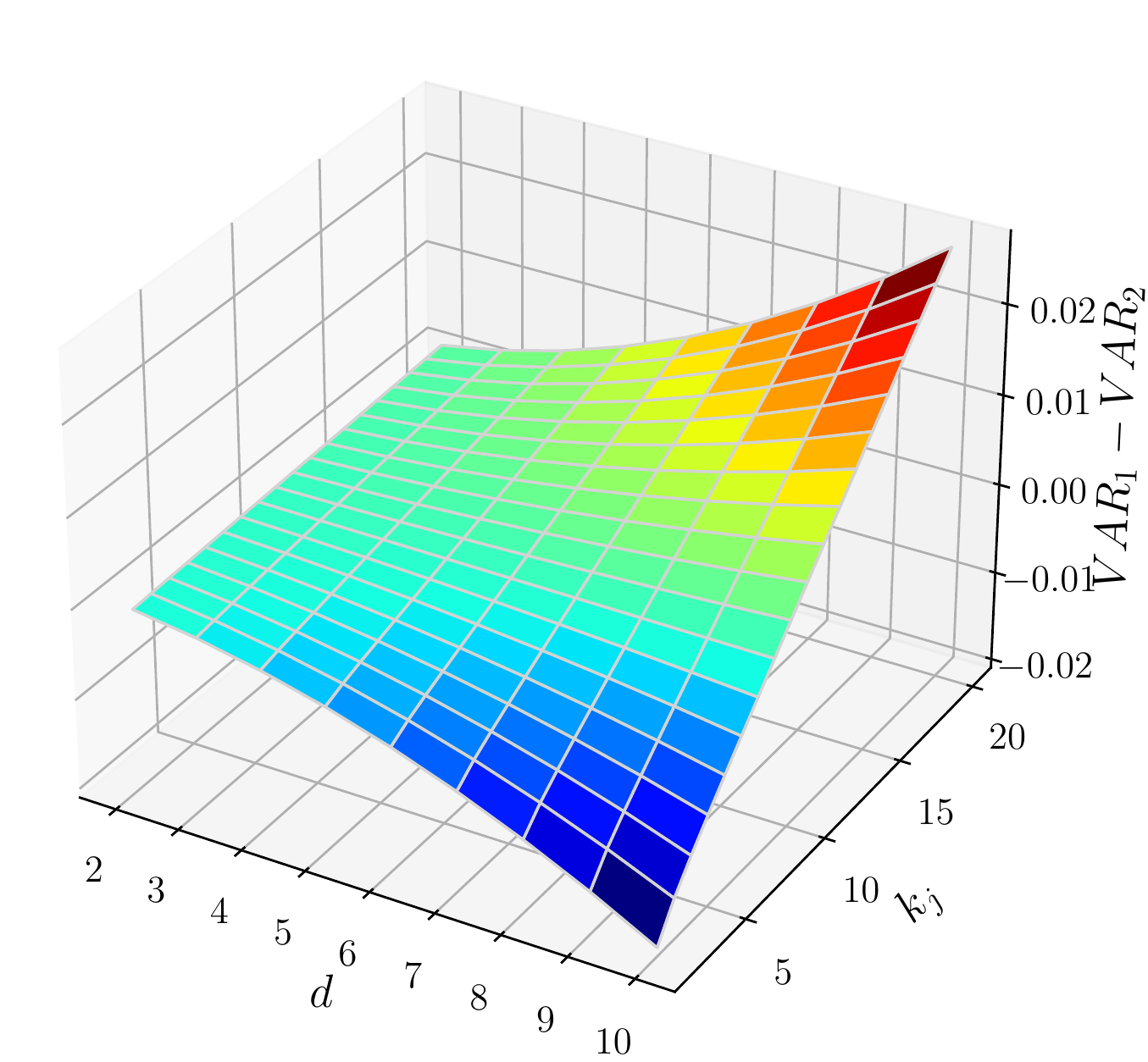}
    \caption{Analytical evaluation of \eqref{ineq:variance} that allows a dynamic selection between RS+FD[GRR] with variance $\operatorname{VAR}_1$ and RS+FD[OUE-z] with variance $\operatorname{VAR}_2$. Parameters were set as $\epsilon'=\ln(3)$, $n=20000$, $d\in[2,10]$, and $k_j\in[2,20]$.}
    \label{fig:surface_variance}
\end{figure}

For the sake of illustration, Fig.~\ref{fig:surface_variance} illustrates a 3D visualization of \eqref{ineq:variance} by fixing $\epsilon'=\ln(3)$ and $n=20000$, and by varying $d\in[2, 10]$ and $k_j \in [2, 20]$, which are common values for real-world datasets (cf. Subsection~\ref{sub:setup}). In this case, one can notice in Fig.~\ref{fig:surface_variance} that neither RS+FD[GRR] nor RS+FD[OUE-z] will always provide the lowest variance value, which reinforces the need for an adaptive mechanism. For instance, with the selected parameters, for lower values of $k_j$, RS+FD[GRR] can provide lower estimation errors even if $d$ is large. On the other hand, as soon as the domain size starts to grow, e.g., $k_j\geq 10$, one is better off with RS+FD[OUE-z] even for small values of $d\geq 3$, as its variance in~\eqref{var:rs+fd_oue_z} does not depend on $k_j$.

\section{Experimental Validation} \label{sec:results}

In this section, we present the setup of our experiments in Subsection~\ref{sub:setup}, the results with synthetic data in Subsection~\ref{sub:synthetic_data}, and the results with real-world data in Subsection~\ref{sub:real_data}.

\subsection{Setup of experiments} \label{sub:setup}

\noindent \textbf{Environment.} All algorithms were implemented in Python 3.8.5 with NumPy 1.19.5 and Numba 0.53.1 libraries. The codes we developed and used for all experiments are available in a Github repository\footnote{\url{https://github.com/hharcolezi/ldp-protocols-mobility-cdrs}}. In all experiments, we report average results over 100 runs as LDP algorithms are randomized.

\noindent \textbf{Synthetic datasets.} Our first set of experiments are conducted on six synthetic datasets. The distribution of values in each attribute follows an uniform distribution, for all synthetic datasets.

\begin{itemize}
    \item For the first two synthetic datasets, we fix the number of attributes $d=5$ and the domain size of each attribute as $\textbf{k}=[10,10,...,10]$ (uniform), and vary the number of users as $n=50000$ and $n=500000$.
    \item Similarly, for the third and fourth synthetic datasets, we fix the number of attributes $d=10$ and the domain size of each attribute as $\textbf{k}=[10,10,...,10]$ (uniform), and vary the number of users as $n=50000$ and $n=500000$.
    \item Lastly, for the fifth and sixth synthetic datasets, we fix the number of users as $n=500000$. Next, we set the number of attributes $d=10$ with domain size of each attribute as $\textbf{k}=[10,20,...,90,100]$ for one dataset, and we set the number of attributes $d=20$ with domain size of each attribute as $\textbf{k}=[10,10,20,20,...,100,100]$ for the other.
\end{itemize}

\noindent \textbf{Real-world datasets.} In addition, we also conduct experiments on four real-world datasets with non-uniform distributions.

\begin{itemize}
    \item \textit{Nursery.} A dataset from the UCI machine learning repository~\cite{uci} with $d=9$ categorical attributes and $n=12960$ samples. The domain size of each attribute is $\textbf{k}=[3, 5, 4, 4, 3, 2, 3, 3, 5]$, respectively. 
    
    \item \textit{Adult.} A dataset from the UCI machine learning repository~\cite{uci} with $d=9$ categorical attributes and $n=45222$ samples after cleaning the data. The domain size of each attribute is $\textbf{k}=[7, 16, 7, 14, 6, 5, 2, 41, 2]$, respectively. 
    
    \item \textit{MS-FIMU.} An open dataset from~\cite{ms_fimu} with $d=6$ categorical attributes and $n=88935$ samples. The domain size of each attribute is $\textbf{k}=[3, 3, 8, 12, 37, 11]$, respectively.
    
    \item \textit{Census-Income.} A dataset from the UCI machine learning repository~\cite{uci} with $d=33$ categorical attributes and $n=299285$ samples. The domain size of each attribute is \begin{math}\textbf{k}=[9, 52, 47, 17,  3,  ..., 43, 43, 43,  5,  3,  3,  3,  2]\end{math}, respectively. 

\end{itemize}

\noindent \textbf{Evaluation and metrics.} We vary the privacy parameter in a logarithmic range as $\epsilon=[\ln (2),\ln (3),...,\ln (7)]$, which is within range of values experimented in the literature for multidimensional data (e.g., in~\cite{wang2019} the range is $\epsilon=[0.5,...,4]$ and in~\cite{Wang2021_b} the range is $\epsilon=[0.1,...,10]$). 

We use the MSE metric averaged per the number of attributes $d$ to evaluate our results. Thus, for each attribute $j$, we compute for each value $v(i) \in \mathcal{D}_j$ the estimated frequency $\hat{f}(v_i)$ and the real one $f(v_i)$ and calculate their differences. More precisely,

\begin{equation}
    MSE_{avg} = \frac{1}{d} \sum_{j \in [1,d]} \frac{1}{|\mathcal{D}_j|} \sum_{v \in \mathcal{D}_j}(f(v_i) - \hat{f}(v_i) )^2 \textrm{.}
\end{equation}

\noindent \textbf{Methods evaluated.} We consider for evaluation the following solutions (cf. Fig.~\ref{fig:system_overview}) and protocols: 

\begin{itemize}

    \item Solution \textit{Spl}, which splits the privacy budget per attribute $\epsilon/d$ with a best-effort approach using the adaptive mechanism presented in Subsection~\ref{sub:adp}, i.e., Spl[ADP]. 
    
    \item Solution \textit{Smp}, which randomly samples a single attribute and use all the privacy budget $\epsilon$ also with the adaptive mechanism, i.e., Smp[ADP].
    
    \item Our solution RS+FD, which randomly samples a single attribute and uses an amplified privacy budget $\epsilon' \geq \epsilon$ while generating fake data for each $d-1$ non-sampled attribute: 
    
    \begin{itemize}
        \item RS+FD[GRR] (Alg.~\ref{alg:rs+fd} with GRR as local randomizer $\mathcal{A}$);
        \item RS+FD[OUE-z] (Alg.~\ref{alg:rs+fd_oue_z});
        \item RS+FD[OUE-r] (Alg.~\ref{alg:rs+fd_oue_r});
        \item RS+FD[ADP] presented in Subsection~\ref{sub:rs+fd_adp} (i.e., adaptive choice between RS+FD[GRR] and RS+FD[OUE-z]).
    \end{itemize}

\end{itemize}

\subsection{Results on synthetic data} \label{sub:synthetic_data}

\begin{figure}[tb]
    \begin{minipage}{1.0\columnwidth}
        \centering
        \includegraphics[width=1\linewidth]{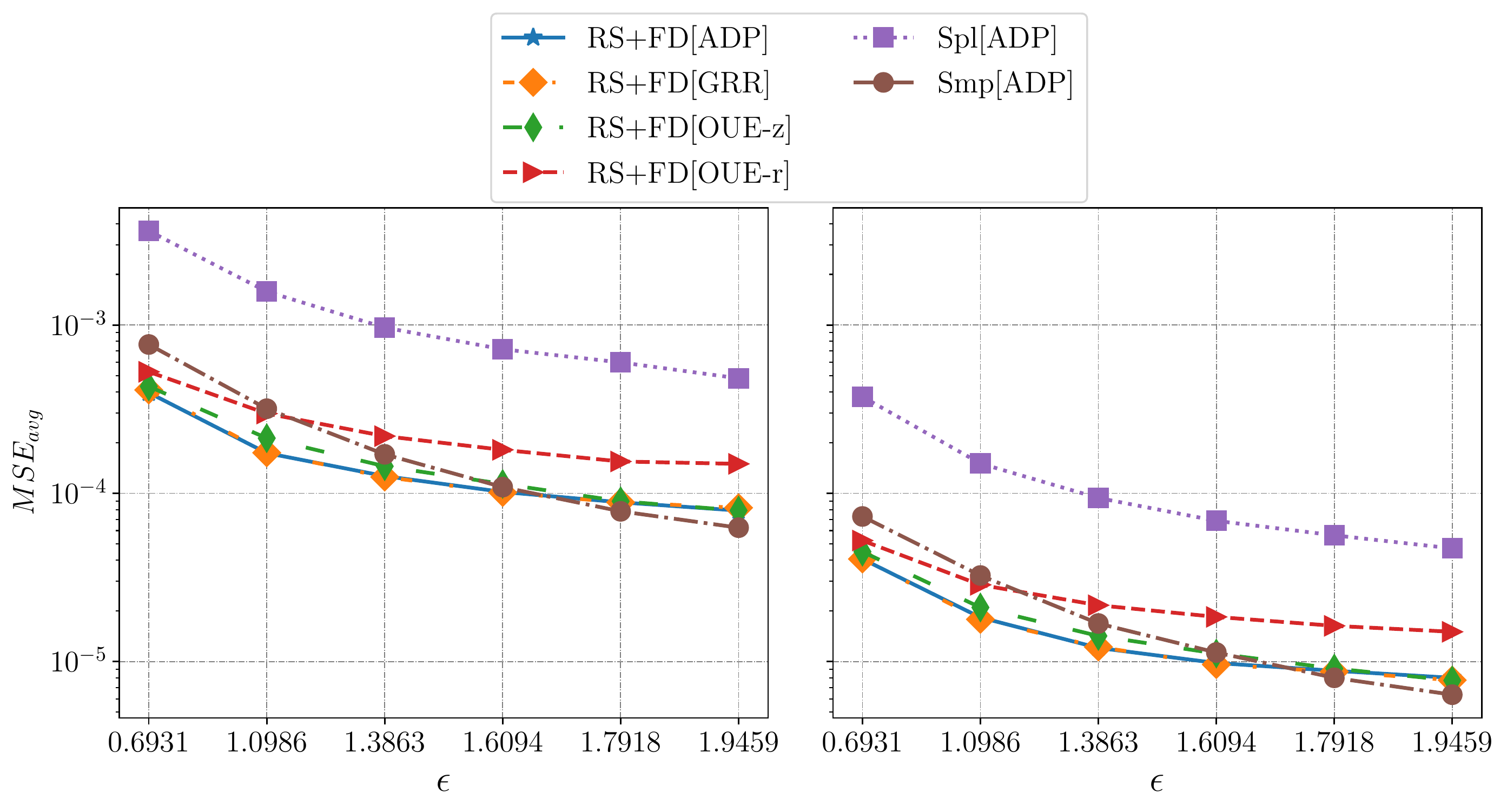}
        \caption{Averaged MSE varying $\epsilon$ on the \textit{synthetic} datasets with $d=5$, uniform domain size $\textbf{k}=[10,10,...,10]$, and $n=50000$ (left-side plot) and $n=500000$ (right-side plot).}\label{fig:results_syn1_syn2}
    \end{minipage}
    \\
    \begin{minipage}{1.0\columnwidth}
        \centering
        \includegraphics[width=1\linewidth]{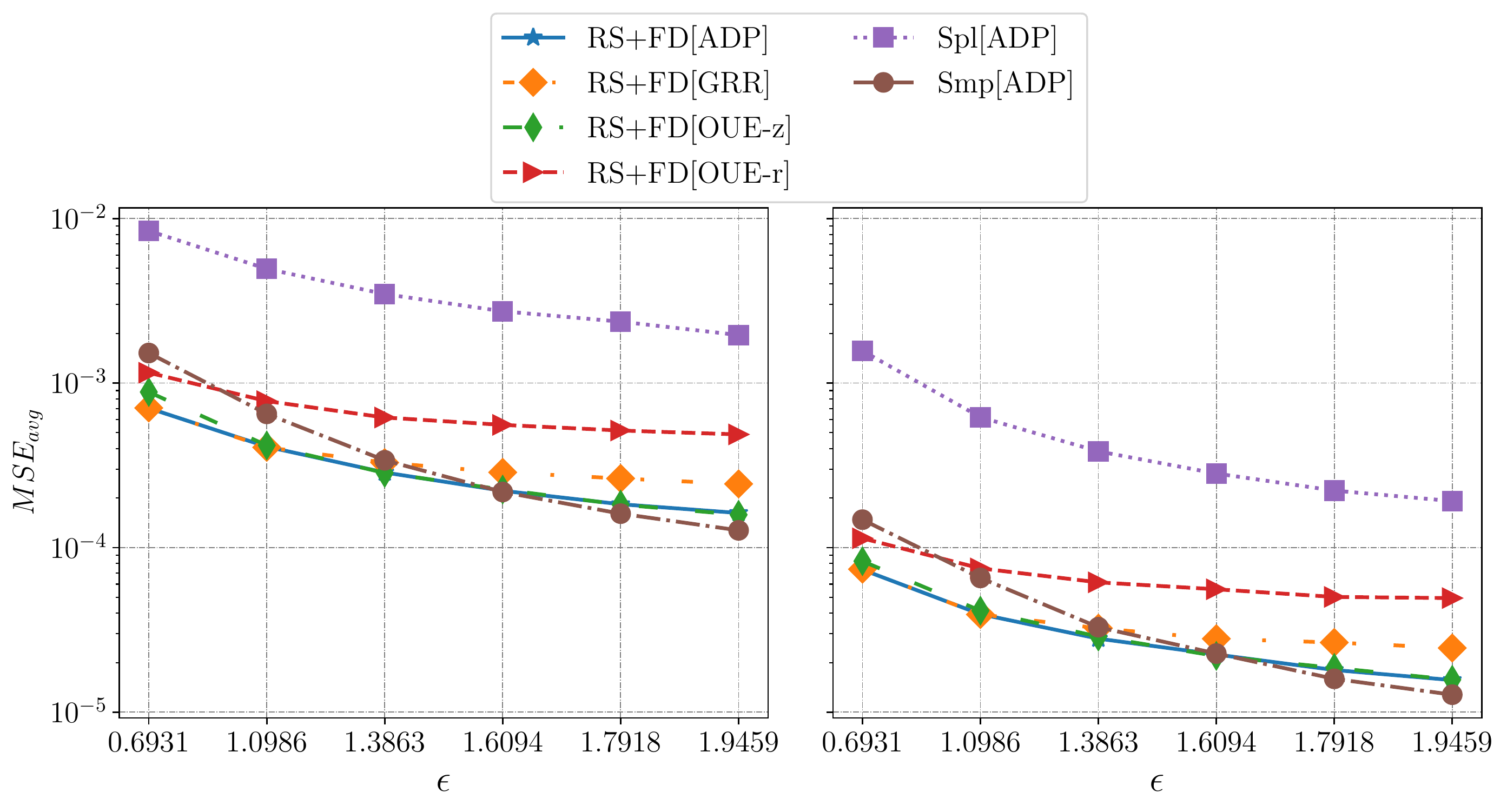}
        \caption{Averaged MSE varying $\epsilon$ on the \textit{synthetic} datasets with $d=10$, uniform domain size $\textbf{k}=[10,10,...,10]$, and $n=50000$ (left-side plot) and $n=500000$ (right-side plot).}\label{fig:results_syn3_syn4}
    \end{minipage}\\
    \begin{minipage}{1.0\columnwidth}
        \centering
        \includegraphics[width=1\linewidth]{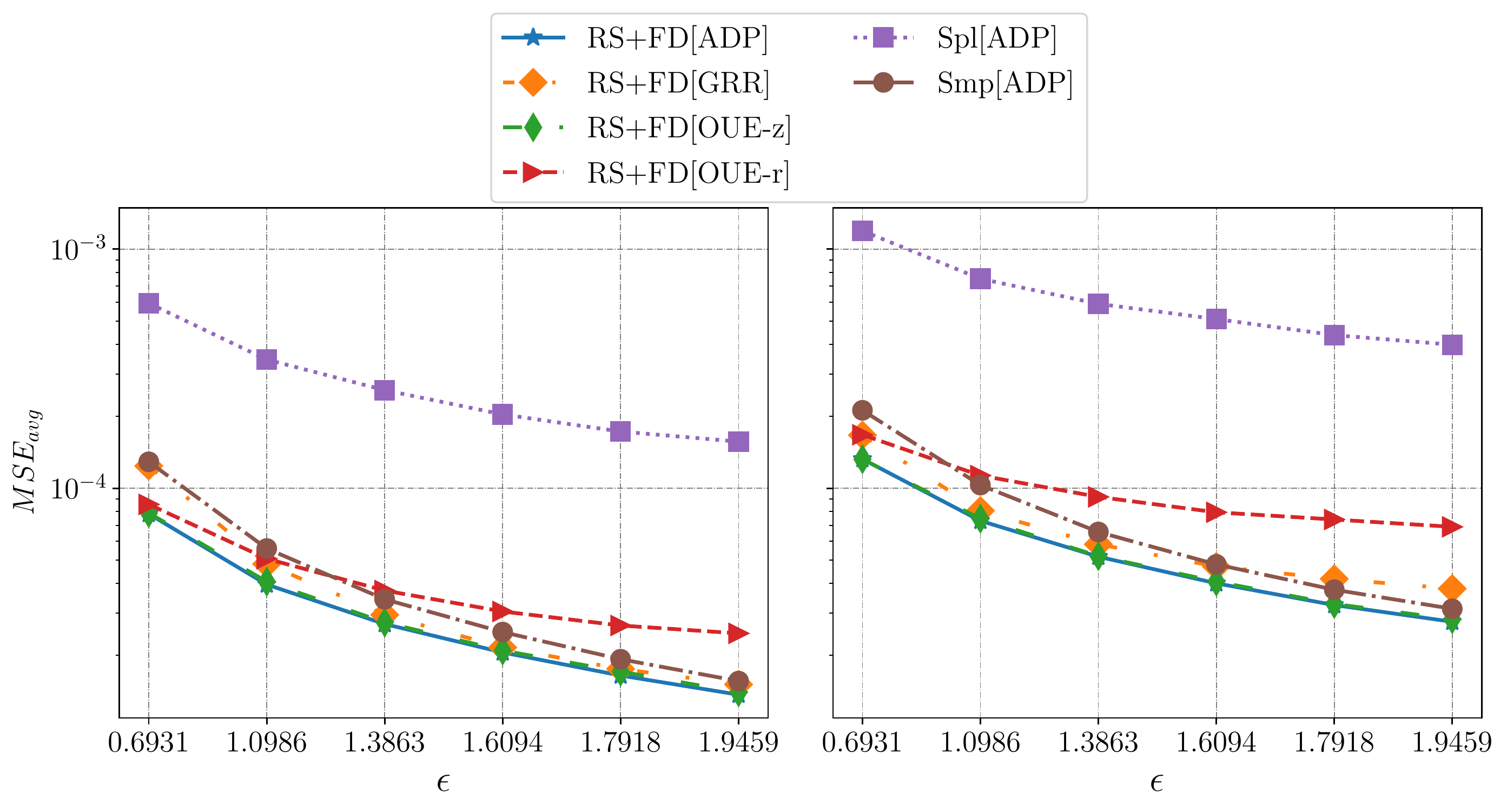}
        \caption{Averaged MSE varying $\epsilon$ on the \textit{synthetic} datasets with $n=500000$: the first with $d=10$ and domain size $\textbf{k}=[10,20,...,90,100]$ (left-side plot), and the other with $d=20$ and domain size $\textbf{k}=[10,10,20,...,100,100]$ (right-side plot).}\label{fig:results_syn5_syn6}
    \end{minipage}
\end{figure}

Our first set of experiments were conducted on six synthetic datasets. Fig.~\ref{fig:results_syn1_syn2} (first two synthetic datsets), Fig.~\ref{fig:results_syn3_syn4} (third and fourth synthetic datsets), and Fig.~\ref{fig:results_syn5_syn6} (last two synthetic datasets) illustrate for all methods, the averaged $MSE_{avg}$ (y-axis) according to the privacy parameter $\epsilon$ (x-axis). 

\noindent \textbf{Impact of the number of users.} In both Fig.~\ref{fig:results_syn1_syn2} and Fig.~\ref{fig:results_syn3_syn4}, one can notice that the $MSE_{avg}$ is inversely proportional to the number of users $n$. With the datasets we experimented, the $MSE_{avg}$ decreases one order of magnitude by increasing $n$ in one order of magnitude too. In comparison with \textit{Smp}, the noise in our RS+FD solution comes mainly from fake data as it uses an amplified $\epsilon' \geq \epsilon$. \textit{This suggests that, in some cases, with appropriately high number of user $n$, our solutions may always provide higher data utility than the state-of-the-art \textit{Smp} solution (e.g., cf. Fig.~\ref{fig:results_syn5_syn6}).}

\noindent \textbf{Impact of the number of attributes.} One can notice the effect on increasing $d$ comparing the results of Fig.~\ref{fig:results_syn1_syn2} ($d=5$) and Fig.~\ref{fig:results_syn3_syn4} ($d=10$) while fixing $n$ and $\textbf{k}$ (uniform number of values). For instance, even though there are twice the number of attributes, the accuracy (measured with the averaged MSE metric) does not suffer much. \textit{This is because the amplification by sampling ($\frac{e^{\epsilon'}-1}{e^{\epsilon}-1} = \frac{1}{\beta}$~\cite{Li2012}) depends on the sampling rate $\beta=\frac{1}{d}$, which means that the more attributes one collects, the more the $\epsilon'$ is amplified, i.e., $\epsilon'=\ln{\left( d \cdot (e^{\epsilon} - 1) + 1 \right)}$; thus balancing data utility. }

Besides, in Fig.~\ref{fig:results_syn5_syn6}, one can notice a similar pattern, i.e., increasing the number of attributes from $d=10$ (left-side plot) to $d=20$ (right-hand plot), with varied domain size $\textbf{k}$, resulted in only a slightly loss of performance. This, however, is not true for the \textit{Spl} solution, for example, in which the $MSE_{avg}$ increased much more in order of magnitude than our RS+FD solution.

\noindent \textbf{Comparison with existing solutions.} From our experiments, one can notice that the  \textit{Spl} solution always resulted in more estimation error than our RS+FD solution and than the \textit{Smp} solution, which is in accordance with other works~\cite{wang2019,xiao2,tianhao2017,Arcolezi2021,Wang2021_b}. Besides, our RS+FD[GRR], RS+FD[OUE-z], and RS+FD[ADP] protocols achieve better or nearly the same performance than the \textit{Smp} solution with a best-effort adaptive mechanism Smp[ADP], which uses GRR for small domain sizes $k$ and OUE for large ones. Although this is not true with RS+FD[OUE-r], it still provides more accurate results than Spl[ADP] while "hiding" the sampled attribute from the aggregator.

\textbf{Globally, on high privacy regimes (i.e., low values of $\epsilon$), our RS+FD solution consistently outperforms the other two solutions \textit{Spl} and \textit{Smp}. By increasing $\epsilon$, Smp[ADP] starts to outperform RS+FD[OUE-r] while achieving similar performance than our RS+FD[GRR], RS+FD[OUE-z], and RS+FD[ADP] solutions.} In addition, one can notice in Fig.~\ref{fig:results_syn3_syn4}, for example, the advantage of RS+FD[ADP] over our protocols RS+FD[GRR] and RS+FD[OUE-z] applied individually, as it adaptively selects the protocol with the smallest \textit{approximate variance} value.

\subsection{Results on real world data} \label{sub:real_data}

\begin{figure*}[t]
    \begin{minipage}[l]{1.0\columnwidth}
        \centering
        \includegraphics[width=0.8\linewidth]{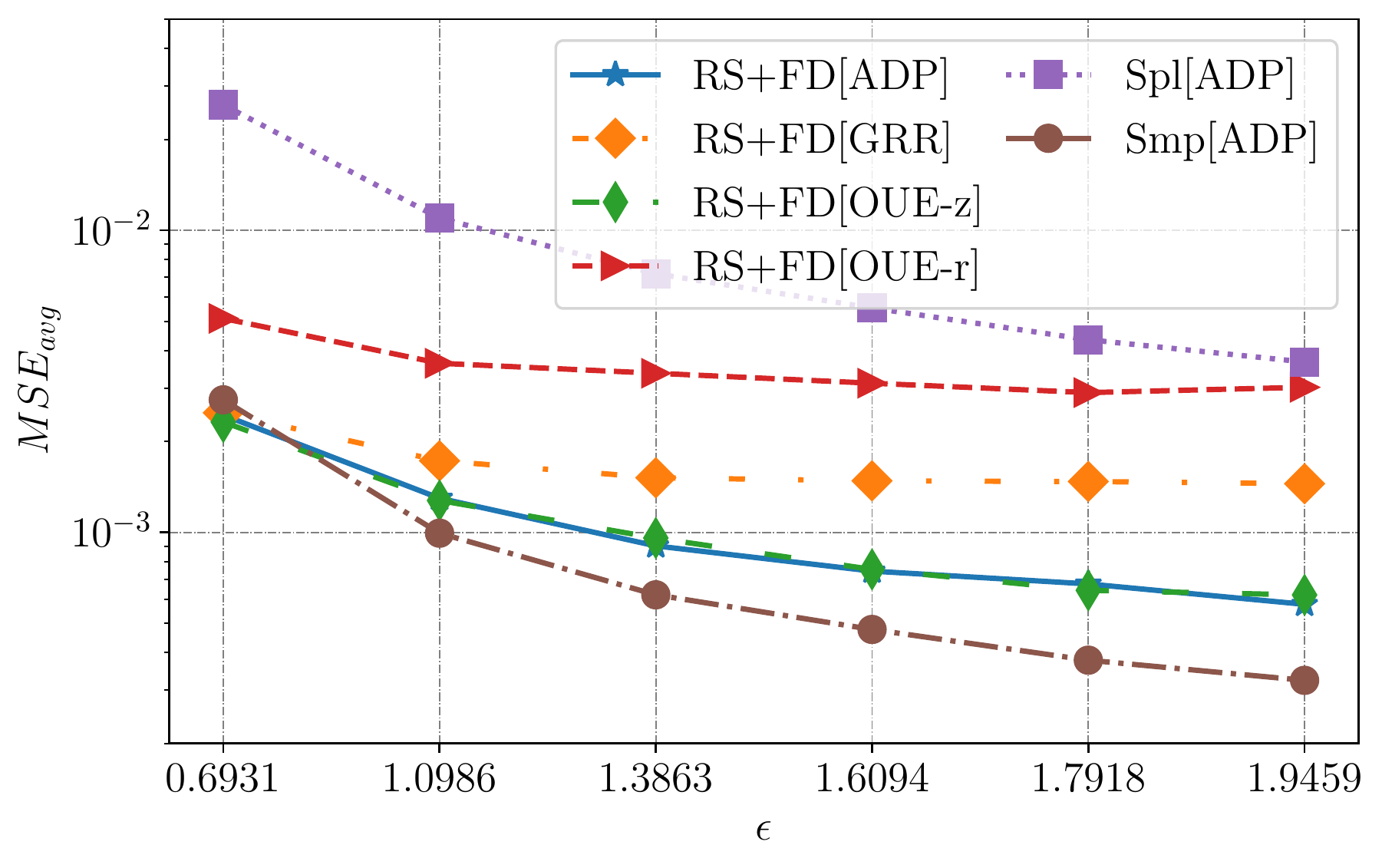}
        \caption{Averaged MSE varying $\epsilon$ on the \textit{Nursery} dataset with $n=12960$, $d=9$, and domain size $\textbf{k}=[3, 5, 4, 4, 3, 2, 3, 3, 5]$.}\label{fig:results_nursery}
    \end{minipage}
    \hfill{}
    \begin{minipage}[r]{1.0\columnwidth}
        \centering
        \includegraphics[width=0.8\linewidth]{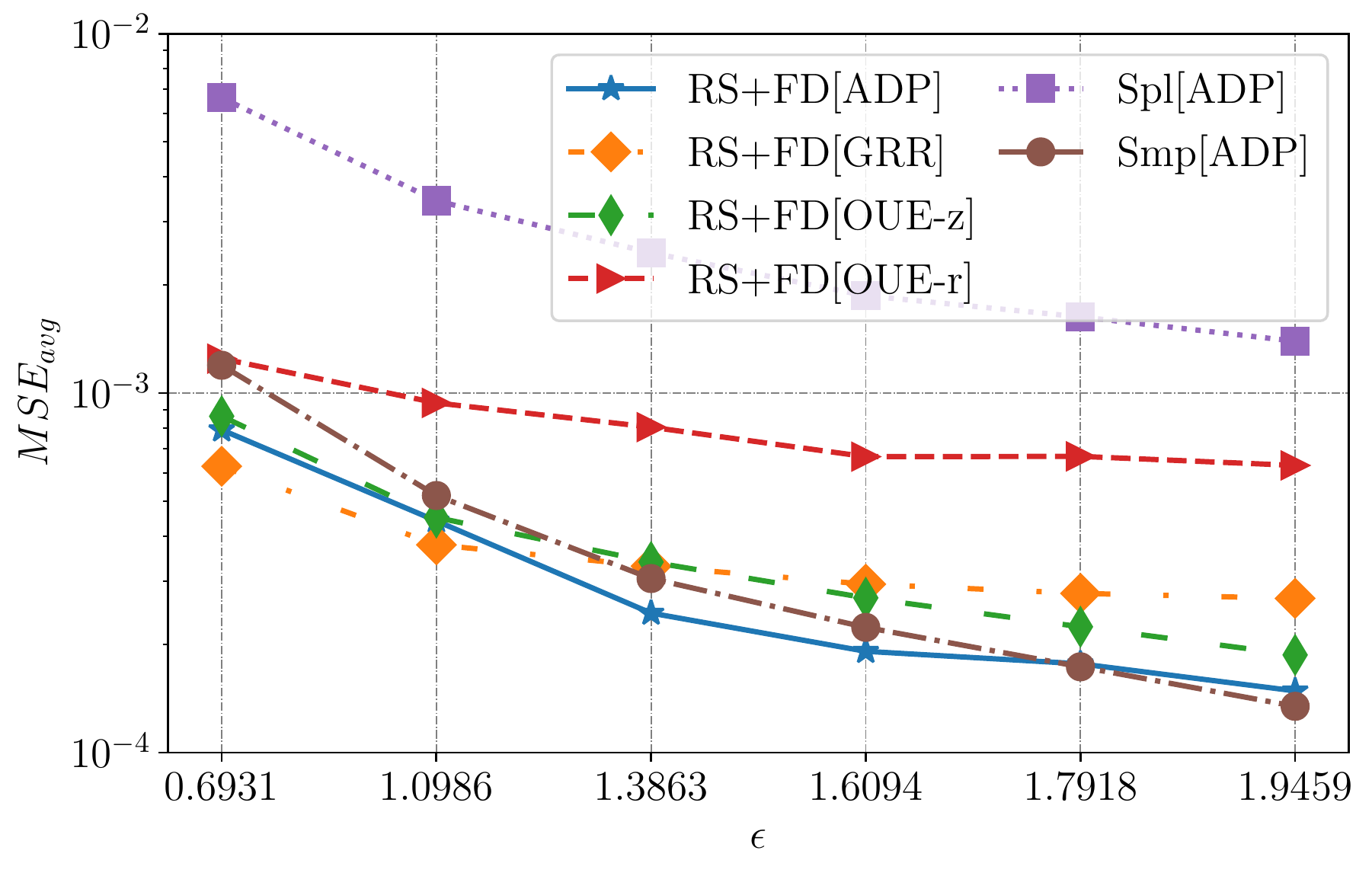}
        \caption{Averaged MSE varying $\epsilon$ on the \textit{Adult} dataset with $n=45222$, $d=9$, and domain size $\textbf{k}=[7, 16, 7, 14, 6, 5, 2, 41, 2]$.}\label{fig:results_adults}
    \end{minipage}\\
    \begin{minipage}[l]{1.0\columnwidth}
        \centering
        \includegraphics[width=0.8\linewidth]{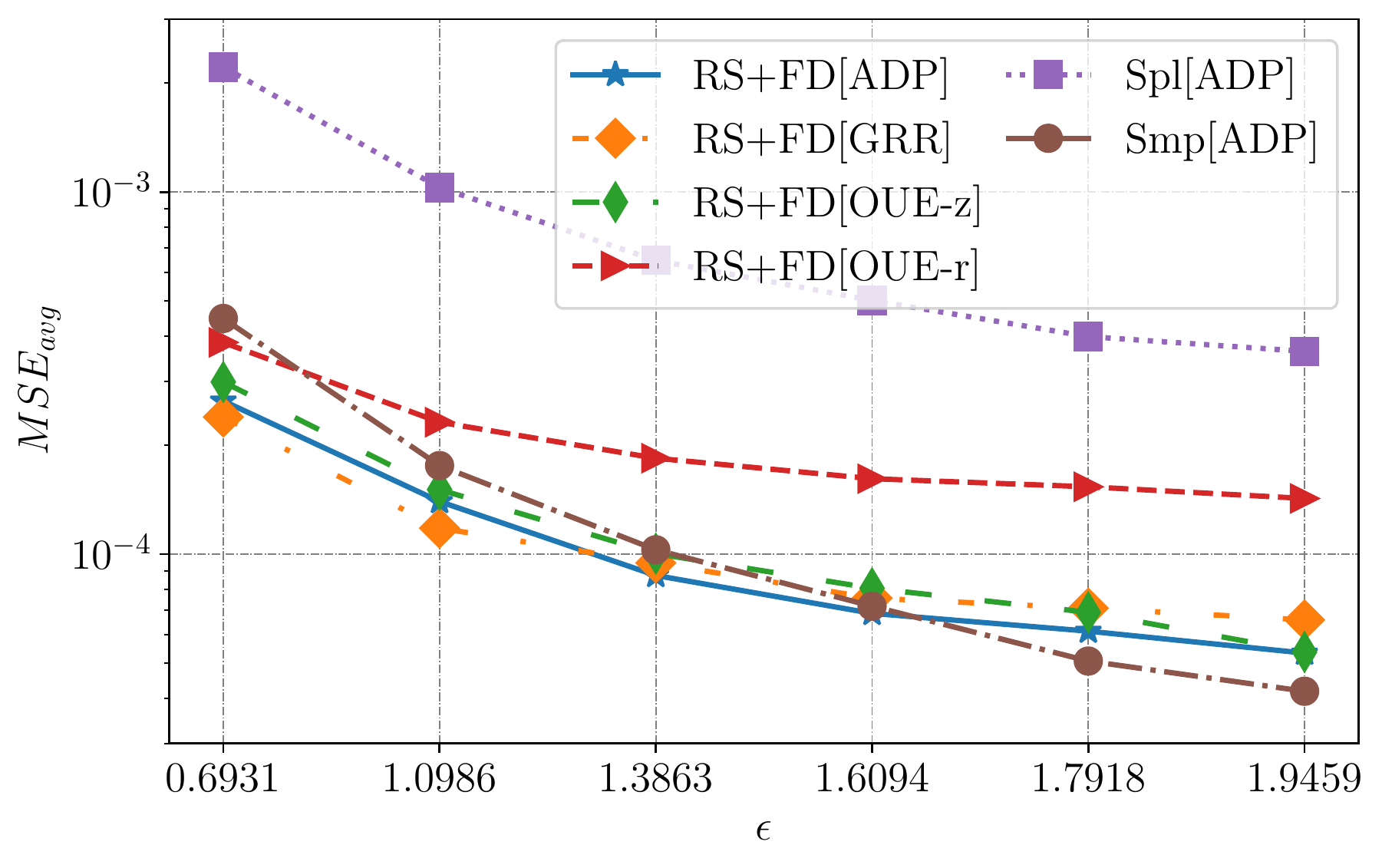}
        \caption{Averaged MSE varying $\epsilon$ on the \textit{MS-FIMU} dataset with $n=88935$, $d=6$, and domain size $\textbf{k}=[3, 3, 8, 12, 37, 11]$.}\label{fig:results_vhs}
    \end{minipage}
    \hfill{}
    \begin{minipage}[r]{1.0\columnwidth}
        \centering
        \includegraphics[width=0.8\linewidth]{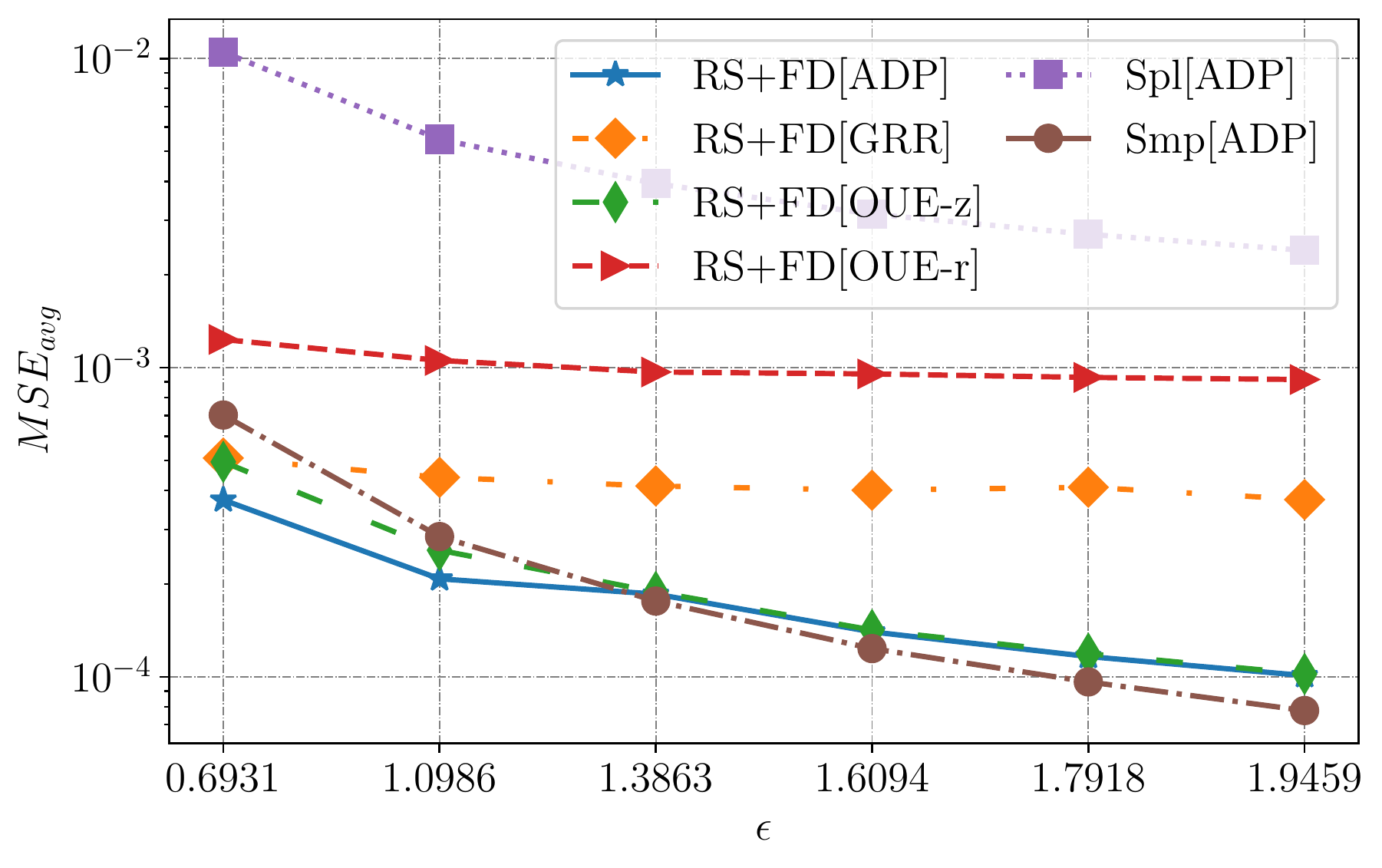}
        \caption{Averaged MSE varying $\epsilon$ on the \textit{Census-Income} dataset with $n=299285$, $d=33$, and domain size \begin{math}\textbf{k}=[9, 52, 47, 17,  3,  ..., 43, 43, 43,  5,  3,  3,  3,  2]\end{math}.}\label{fig:results_census}
    \end{minipage}
\end{figure*}

Our second set of experiments were conducted on four real-world datasets with varied parameters for $n$, $d$, and $\textbf{k}$. Fig.~\ref{fig:results_nursery} (\textit{Nursery}), Fig.~\ref{fig:results_adults} (\textit{Adult}), Fig.~\ref{fig:results_vhs} (\textit{MS-FIMU}), and Fig.~\ref{fig:results_census} (\textit{Census-Income}) illustrate for all methods, averaged $MSE_{avg}$ (y-axis) according to the privacy parameter $\epsilon$ (x-axis).

The results with real-world datasets follow similar behavior than with synthetic ones. For all tested datasets, one can observe that the $MSE_{avg}$ of our proposed protocols with RS+FD is still smaller than the \textit{Spl} solution with a best-effort adaptive mechanism Spl[ADP]. As also highlighted in the literature~\cite{wang2019,xiao2,tianhao2017,Arcolezi2021,Wang2021_b}, privacy budget splitting is sub-optimal, which leads to higher estimation error.  

On the other hand, for both \textit{Adult} and \textit{MS-FIMU} datasets, our solutions RS+FD[GRR], RS+FD[OUE-z], and RS+FD[ADP] achieve nearly the same performance (sometimes better on high privacy regimes, i.e., low $\epsilon$) than the \textit{Smp} solution with the best-effort adaptive mechanism Smp[ADP]. For the \textit{Nursery} dataset, with small number of users $n$, only RS+FD[OUE-z] and RS+FD[ADP] are competitive with Smp[ADP]. Lastly, for the \textit{Census} dataset, with a large number of attributes $d=33$, increasing the privacy parameter $\epsilon$ resulted in a small gain on data utility for our solutions RS+FD[GRR] and RS+FD[OUE-r]. On the other hand, both of our solutions RS+FD[OUE-z] and RS+FD[ADP] achieve nearly the same or better performance than Smp[ADP]. 

Moreover, one can notice that using the \textit{approximate variance} in~\eqref{ineq:variance} led RS+FD[ADP] to achieve similar or improved performance over our RS+FD[GRR] and RS+FD[OUE-z] protocols applied individually. For instance, for the \textit{Adult} dataset, with RS+FD[ADP] it was possible to outperform Smp[ADP] 3x more than with RS+FD[GRR] or RS+FD[OUE-z] (similarly, 1x more for the \textit{MS-FIMU} dataset). Besides, for the \textit{Census-Income} dataset, RS+FD[ADP] improves the performance of the other protocols applied individually on high privacy regimes while accompanying the RS+FD[OUE-z] curve on the lower privacy regime cases.

\textbf{In general, these results help us answering the problematic of this paper (cf. Subsection~\ref{sub:problematic}) that for the same privacy parameter $\epsilon$, one can achieve nearly the same or better data utility with our RS+FD solution than when using the state-of-the-art \textit{Smp} solution. Besides, RS+FD enhances users' privacy by "hiding" the sampled attribute and its $\epsilon$-LDP value among fake data.} On the other hand, there is a price to pay on computation, in the generation of fake data, and on communication cost, which is similar to the \textit{Spl} solution, i.e., send a value per attribute.

\section{Literature Review and Discussion} \label{sec:discussion}

In recent times, there have been several works on the local DP setting in both academia~\cite{wang2019,xiao2,first_ldp,tianhao2017,Alvim2018,Hadamard,Bassily2015,Xiong2020,kairouz2016discrete,Wang2021_b} and practical deployment~\cite{rappor,microsoft,apple,Kessler2019}. Among many other complex tasks (e.g., heavy hitter estimation~\cite{Bassily2015,Wang2021,bassily2017practical}, marginal estimation~\cite{Peng2019,Zhang2018,Ren2018,Fanti2016}, frequent itemset mining~\cite{Wang2018,Qin2016}), frequency estimation is a fundamental primitive in LDP and has received considerable attention for a single attribute~\cite{Hadamard,Wang2021_b,tianhao2017,kairouz2016discrete,Alvim2018,rappor,microsoft,Xiong2020,Kim2018,Arcolezi2020,Zhao2019,Li2020}. However, concerning multiple attributes, as also noticed in the survey work on LDP in~\cite{Xiong2020}, most studies for collecting multidimensional data with LDP mainly focused on numerical data~\cite{xiao2,Duchi2018,wang2019,Wang2021_b}. For instance, in~\cite{xiao2,wang2019}, the authors propose sampling-based LDP mechanisms for real-valued data (named Harmony and Piecewise Mechanism) and applied these protocols in a multidimensional setting using state-of-the-art LDP mechanisms from~\cite{Bassily2015,tianhao2017} for categorical data. On the other hand, regarding categorical attributes, in~\cite{tianhao2017}, the authors prove for OUE (as well as to optimal local hashing - OLH) that sending $1$ attribute with the whole privacy budget $\epsilon$ results in less variance than splitting the privacy budget $\epsilon/d$ for all attributes. The authors in~\cite{Arcolezi2021} prove and validate experimentally that reporting a single attribute with $\epsilon$-LDP resulted in less estimation error than splitting the privacy budget when using GRR.

However, in the aforementioned works~\cite{xiao2,wang2019,tianhao2017,Arcolezi2021,Wang2021_b}, the sampling result is known by the aggregator. That is, each user samples a single attribute $j$, applies a local randomizer to $v_j$, and sends to the aggregator the tuple $y=\langle j, LDP(v_j) \rangle$ (i.e., \textit{Smp}). While one can achieve higher data utility (cf. Figs.~\ref{fig:results_syn1_syn2}-~\ref{fig:results_census}) with \textit{Smp} than splitting the privacy budget among $d$ attributes (\textit{Spl}), we argue that \textit{Smp} might be "unfair" with some users. More precisely, users whose sampled attribute is socially "more" sensitive (e.g., disease or location), might hesitate to share their data as the probability bound $e^{\epsilon}$ is "less" restrictive than $e^{\epsilon/d}$. For instance, assume that GRR is used with k=2 (HIV positive or negative) and the privacy budget is $\epsilon=ln(7) \sim 2$, the user will report the true value with probability as high as $p \sim 87\%$ (even with $\epsilon=1$, this probability is still high $p \sim 73\%$). On the other hand, if there are $d=10$ attributes (e.g., nine demographic and HIV test), with \textit{Spl}, the probability bound is now $e^{\epsilon/10}$ and $p \sim 55\%$.

Motivated by this privacy-utility trade-off between the solutions \textit{Spl} and \textit{Smp}, we proposed a solution named random sampling plus fake data (RS+FD), which generates uncertainty over the sampled attribute in the view of the aggregator. In this context, since the sampling step randomly selects an attribute with sampling probability $\beta=\frac{1}{d}$, there is an amplification effect in terms of privacy, a.k.a. amplification by sampling~\cite{Chaudhuri2006,Li2012,balle2018privacy,balle2020privacy,first_ldp}. A similar privacy amplification for sampling a random item of a single attribute has been noticed in~\cite{Wang2018} for frequent itemset mining in the LDP model too. Indeed, \textit{amplification} is an active research field on DP literature, which aims at finding ways to measure the privacy introduced by non-compositional sources of randomness, e.g., sampling~\cite{Chaudhuri2006,Li2012,balle2018privacy,balle2020privacy,first_ldp}, iteration~\cite{Feldman2018}, and shuffling~\cite{Balle2019,Erlingsson2019,erlingsson2020encode,Wang2020}. 

\section{Conclusion and Perspectives} \label{sec:conclusion}

This paper investigates the problem of collecting multidimensional data under $\epsilon$-LDP for the fundamental task of frequency estimation. As shown in the results, our proposed RS+FD solution achieves nearly the same or better performance than the state-of-the-art \textit{Smp} solution. In addition, RS+FD generates uncertainty over the sampled attribute in the view of the aggregator, which enhances users' privacy. For future work, we suggest and intend to investigate if given a reported tuple $\textbf{y}$ one can state which attribute value is "fake" or not by seeing the estimated frequencies. Indeed, we intend to investigate this phenomenon in both single-time collection and longitudinal studies (i.e., throughout time) by extending RS+FD with two rounds of privatization, i.e., using \textit{memoization}~\cite{rappor,microsoft}.

\begin{acks}
This work was supported by the Region of Bourgogne Franche-Comt\'e CADRAN Project and by the EIPHI-BFC Graduate School (contract ``ANR-17-EURE-0002"). All computations have been performed on the "M\'esocentre de Calcul de Franche-Comt\'e".
\end{acks}

\bibliographystyle{ACM-Reference-Format}
\bibliography{ms.bib}

\end{document}